\documentclass[twocolumn,nofootinbib,amsmath,amssymb,aps,prd,balancelastpage,superscriptaddress]{revtex4-1}

\usepackage{color}
\usepackage[active]{srcltx}
\usepackage{amsmath,amsfonts,amssymb,amsthm,amstext,amscd,eucal,srcltx}
\usepackage{epsfig,graphicx,bm}
\usepackage{epstopdf, epsf}
\usepackage{dcolumn}
\usepackage{hyperref}

\newcommand{\be}{\begin{equation}}
\newcommand{\ee}{\end{equation}}

\newcommand{\bse}{\begin{subequations}}
\newcommand{\ese}{\end{subequations}}
\newcommand{\bea}{\begin{eqnarray}}
\newcommand{\eea}{\end{eqnarray}}
\newcommand{\ba}{\begin{array}}
\newcommand{\ea}{\end{array}}
\newcommand{\bc}{\begin{center}}
\newcommand{\ec}{\end{center}}

\begin{document}
\preprint{IPM/P-2012/009}  
\vspace*{3mm}

\title{Analogous BMS Symmetry in QED and Quantum Anomaly in Dirac and Weyl Semimetals}%

\author{Andrea Addazi}
\email{andrea.addazi@lngs.infn.it}
\affiliation{Center for Field Theory and Particle Physics \& Department of Physics, Fudan University, 200433 Shanghai, China}

\author{Antonino Marcian\`o}
\email{marciano@fudan.edu.cn}
\affiliation{Center for Field Theory and Particle Physics \& Department of Physics, Fudan University, 200433 Shanghai, China}

\begin{abstract}
\noindent
We show that the asymptotic infinite dimensional enlarged gauge symmetries constructed for QED are anomalous in Weyl semimetals. This symmetry is particularly important in particle physics for its analogy with the Bondi-Metzner-Sachs (BMS) symmetries in gravity, as well as for its connection with QED soft IR theorems. This leads to observable effects, because of the induction of a new current in the material, which carries a memory of the BMS symmetry precursor. 
\end{abstract}

\maketitle

\noindent
{\it Introduction. --- } Dirac and Weyl semimetals are an exciting {\it test-bed} for many important concepts and frameworks of quantum field theory. Indeed, it was argued very recently that the chiral currents generated in these materials are understood as a manifestation of a chiral anomaly term. An holographic explanation of the phenomena was found \cite{Gooth:2017mbd,Landsteiner:2017lwm,Landsteiner:2017hye}. This was developed connecting the quantum field theory description of the chiral anomalous phenomena to the anomalous terms of (an effective) string theory in higher dimensional anti-de Sitter bulk. 
Quantum anomalies are important in our understanding of high energy physics and in particular of the Standard Model of particles and interactions. As it is well known, the chiral anomaly induced by 1-loop triangles of quarks, explained the pion decay into two photons\footnote{We thank Jerzy Kowalsky-Glikman and Paolo Pani for reminding us to add this remark.}, reproducing the correct amount for the transitions $\pi \rightarrow \gamma\gamma$ observed in the laboratory.

\vspace{2mm}

It is also well known that all gauge anomalies of the Standard Model vanish by virtue of a perfect balance among the contributions arising from all the matter particles. This cancellation appears as an accidental miracle that allows the Standard Model to be not plagued by unitarity and causality loss. Finally, an anomaly of the scale invariance and conformal symmetry is dynamically generated by radiative corrections or by non-perturbative phenomena --- this for instance the case of confinement in Quantum Chromodynamics. About the conformal anomaly, it was recently suggested that it may be tested in the very same Weyl-semimetals \cite{Chu:2018ksb,Chu:2018ntx,Chernodub:2017jcp,Arjona:2019lxz,Chernodub:2019blw}. Indeed, for a boundary plane with a magnetic field parallelly   oriented, the conformal Weyl anomaly induces a charge transport current. 

\vspace{2mm}

In this letter, we propose a new paradigm of testing anomalies of infinite dimensional asymptotic symmetries. The Bondi-Metzner-Sachs (BMS) symmetry is certainly the most popular infinite dimensional asymptotic symmetry of gravity fields. As shown by Strominger and collaborators in a series of papers, BMS is related to the same Ward identities of the soft infrared gravitational radiation \cite{Strominger:2013jfa,He:2014laa,Strominger:2014pwa}. Recently Hawking, Perry and Strominger proposed that the very same same black hole information loss paradox may be solved invoking a memory effect from the new Noether charges associated to BMS \cite{Hawking:2016msc}. BMS symmetry is an extension of the conformal Weyl symmetry with new infinite supertranslation symmetries. Consequently BMS includes an infinite number of supertranslation charges, which may provide a gravitational memory effect. It has been also argued that an analogous of the BMS symmetry is recovered also in Quantum Electrodynamics, as a large gauge symmetry \cite{He:2014cra,Kapec:2017tkm,Henneaux:2018gfi}.

\vspace{2mm}

A large gauge symmetry incorporates an infinite number of gauge symmetries \cite{He:2014cra,Kapec:2017tkm,Henneaux:2018gfi}. Once again, this corresponds to an infinite number of Noether charges, and is exactly the symmetry that we claim to be anomalous in the Weyl semimetals. We unveil the appearance of this quantum anomaly in a simple set-up: a boundary plane with a constant magnetic field parallelly oriented on it. The magnetic field induces, through the analogous BMS anomaly, an electric current which leads to a characteristic potential and charge redistribution. Such an effect is related to the formation of Schwinger fermion pairs, which percolates into a characteristic electric potential and a Debye screening of the electric field. This modifies the standard electrostatic and vector potentials in the asymptotic limit as the breaking of the Large Asymptotic QED symmetry.

\vspace{2mm}

{\it BMS and large asymptotic gauge symmetries. --- } Let us initially start from the standard vacuum theory in Minkowski space-time:
\begin{equation}
\label{Metric}
ds^{2}=-dT^{2}+dZ^{2}+dY^{2}+dX^{2},\quad (c=1).
\end{equation}
We may complexify the variables through the change of coordinates
\begin{equation}
X^{2}+Y^{2}+Z^{2}=r^{2},\qquad T=u+r\, ,
\end{equation}
\begin{equation}
\label{aal}
Z=r\frac{1-z\bar{z}}{1+z \bar{z}}, \qquad X+iY=\frac{2rz}{1+z\bar{z}}\, .
\end{equation}
This allows to obtain the metric of the Minkowski spacetime in the future null infinity $(\mathcal{I}^{+})$, as shown in the Penrose's diagram in Fig.1, namely 
\begin{equation}
\label{dsuu}
ds^{2}=-du^{2}-2dudr+2\gamma_{z\bar{z}}dz d\bar{z}\, , 
\end{equation}
where $u=T-R$ and $\gamma_{z\bar{z}}=2/(1+z\bar{z})^{2}$ is the metric of the conformal sphere. The asymptotic topology is $S^{2}\times R$ at the $r=\infty$ boundary. Boundaries of $\mathcal{I}^{+}$ at $u=\pm \infty$ are denoted as $\mathcal{I}_{\pm}^{+}$.

The bulk equations are as follows 
\begin{equation}
\label{nablaF}
\nabla^{\nu}\mathcal{F}_{\nu\mu}=q\, j_{\mu}\, , 
\end{equation}
where $\mathcal{F}_{\mu\nu}=\partial_{\mu}\mathcal{A}_{\nu}-\partial_{\nu}\mathcal{A}_{\mu}$ and $j_{\mu}$ is the conserved matter where $\nabla^{\mu}j^{\mu}=0$.

The gauge symmetry associated to the equation of motion (EoM) is 
\begin{equation}
\label{Amu}
\delta_{\epsilon}\mathcal{A}_{\mu}=\partial_{\mu}\epsilon\, . 
\end{equation}

We consider the retarded radial gauge, defined by   
\begin{equation}
\label{Aru}
\mathcal{A}_{r}=0,  \qquad    \mathcal{A}_{u}|_{\mathcal{I}^{+}}=0\, . 
\end{equation}

Let us expand the fields around $\mathcal{I}^{+}$. To ensure a finite non-vanishing radiation flux, $\int_{\mathcal{I}^{+}}\mathcal{F}_{u}^{z}\mathcal{F}_{uz}\neq 0$, a $\mathcal{A}_{z}\sim O(1)$ near $\mathcal{I}^{+}$ is required. On the other hand, Eq.(\ref{Aru}) implies that $\mathcal{A}_{u}\sim O(1/r)$ near $\mathcal{I}^{+}$. This implies that free Maxwell fields undergo the expansion
\begin{eqnarray}
\label{aajjj}
&&\mathcal{A}_{z}(r,u,z,\bar{z})=A_{z}(u,z,\bar{z})+\sum_{n=1}^{\infty}\frac{A_{z}^{(n)}(u,z,\bar{z})}{r^{n}}\, , \\
\label{Auzz}
&&\mathcal{A}_{u}(r,u,z,\bar{z})=\frac{1}{r}A_{u}(u,z,\bar{z})+\sum_{n=1}^{\infty}\frac{A_{u}^{(n+1)}(u,z,\bar{z})}{r^{n}}\, .
\end{eqnarray}
The leading terms of the field strength are $\mathcal{F}_{z\bar{z}}=O(1)$, $\mathcal{F}_{ur}=O(r^{-2})$, $\mathcal{F}_{uz}=O(1)$
and $\mathcal{F}_{rz}=O(r^{-2})$. 
At leading order the constraint equation is 
\begin{equation}
\label{gammap}
\gamma_{z\bar{z}}\partial_{u}A_{u}=\partial_{u}(\partial_{z}A_{\bar{z}}+\partial_{\bar{z}}A_{z})+ q \, 
\gamma_{z\bar{z}}J_{u}\, , 
\end{equation}
where 
\begin{equation}
\label{juuu}
J_{u}(u,z,\bar{z})=
\lim_{r\rightarrow \infty}[r^{2}j_{u}(r,u,z,\bar{z})]\, . 
\end{equation}

The gauge conditions for Eq.(\ref{Aru}) leave unfixed the arbitrary function $\epsilon\equiv \epsilon(z,\bar{z})$ on the conformal sphere at $r=\infty$. These are called Large Gauge Transformations, specified by 
\begin{equation}
\label{delta}
\delta_{\epsilon}A_{z}(u,z,\bar{z})=\partial_{z}\epsilon(z,\bar{z})\, . 
\end{equation}
The charges associated to the gauge theory on $\mathcal{I}^{+}$ are 
\begin{equation}
\label{Qaaa}
Q^{+}_{\epsilon^{+}}=\int_{\mathcal{I}_{+}} du d^{2}z \epsilon [\partial_{u}(\partial_{z}A_{\bar{z}}+\partial_{\bar{z}} A_{z})+q\,\gamma_{z\bar{z}}J_{u}]\,.
\end{equation}
A similar expression for the charges can be found in the past boundary $\mathcal{I}^{-}$:
\begin{equation}
\label{kaka}
Q_{\epsilon^{-}}^{-}=\int_{\mathcal{I}_{-}}dv d^{2}z\epsilon^{-}[\partial_{v}(\partial_{z}A_{\bar{z}}+\partial_{\bar{z}} A_{z})+q\,\gamma_{z\bar{z}}J_{v}]\, , 
\end{equation}
where now coefficient functions $A$ depend on the advanced coordinate $v$ in stead of the retarded $u$. 

\begin{figure}[ht]
\centerline{ \includegraphics [width=1.1\columnwidth]{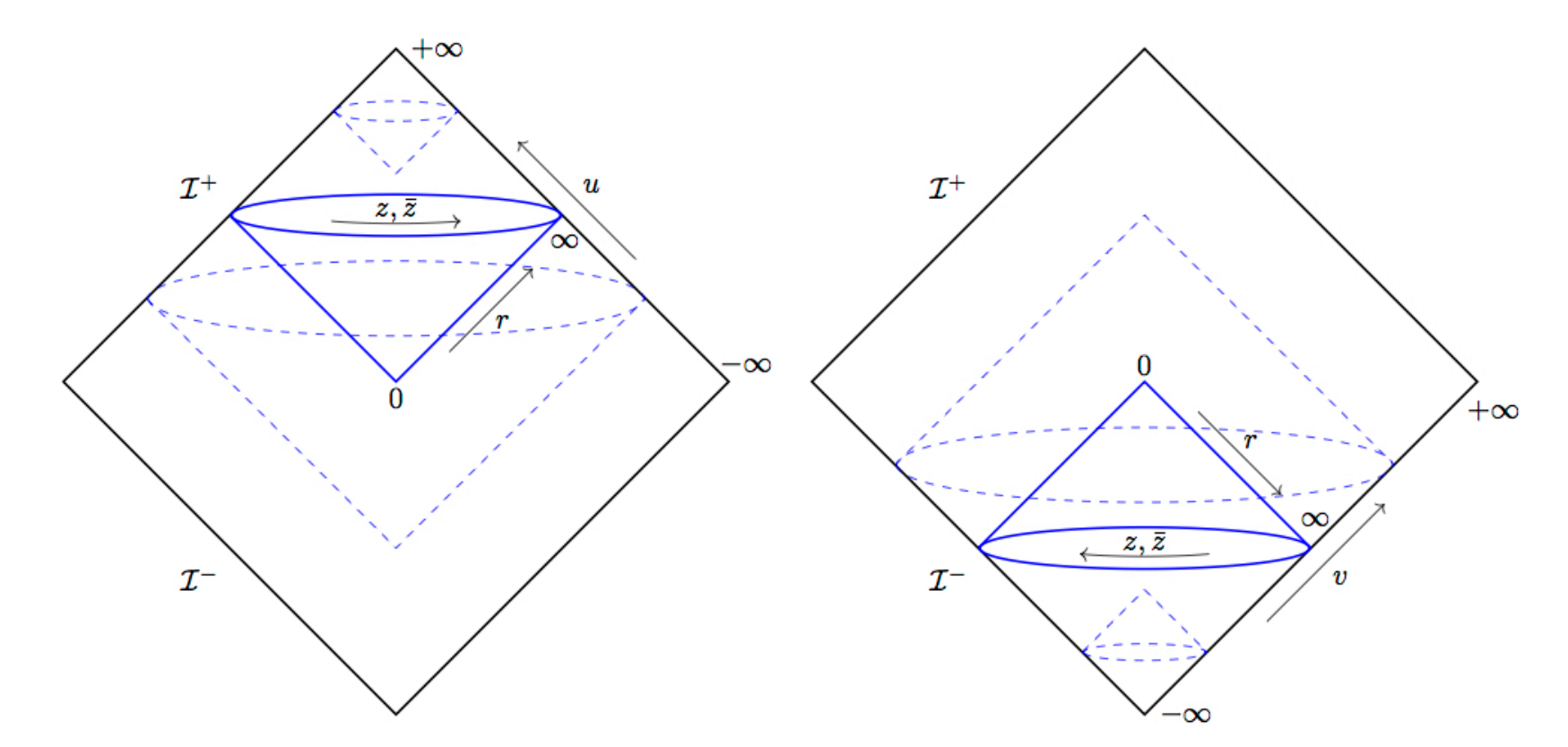}}
\caption{The Penrose diagrams for the future (left panel) and the past (right panel) charts of Minkowski space-time in vacuum are displayed. In these diagrams, in correspondence of every point of the 2D real plane of $r,u(v)$ there is a compactified conformal invariant complex sphere, with $z,\bar{z}$ complex coordinates. $\mathcal{I}^{\pm}$ correspond to the future and past null infinity. The BMS symmetry, as well as the Large Asymptotic Gauge Symmetry, are defined on the $\mathcal{I}^{\pm}$ region lines.}
\end{figure}

The invariance of the theory under the large gauge symmetries generated by the charges casts 
\begin{equation}
\label{ajak}
\langle {\rm OUT} |(Q_{\epsilon}^{+}\mathcal{S}-\mathcal{S}Q_{\epsilon}^{-})|{\rm IN}\rangle=0\, , 
\end{equation}
where generically we have $n$ particles in the in state $|{\rm IN} \rangle=|z_{1}^{\rm IN},...,z_{n}^{\rm IN}\rangle$ and $m$ in OUT as $|{\rm OUT}\rangle=|z_{1}^{\rm OUT},...,z_{m}^{\rm OUT}\rangle$, while $\mathcal{S}$ is the S-matrix operator. 
This implies that for an incoming state
\begin{equation}
\label{Qepsilon}
Q_{\epsilon}^{-}|0_{\rm IN}\rangle=F^{-}[\epsilon]|0_{\rm IN}\rangle\, , 
\end{equation}
which means that the charges do not annihilate the vacuum state, unless $\epsilon$ is fixed to a constant. From this relation, one derives the associated Goldstone bosons, which coincide with the $\phi_{\pm}$ fields entering $A_{z}^{\pm}(z,\bar{z})=\partial_{z}\phi_{\pm}(z,\bar{z})$. The corresponding Ward identities are 
\begin{eqnarray}
\label{Qepep}
&&\langle {\rm OUT} | :F[\epsilon]\, \mathcal{S}:|{\rm IN}\rangle= \\ 
&&\Big[\sum_{k=1}^{n}q_{k}^{\rm IN}\epsilon(z_{k}^{\rm IN},\bar{z}_{k}^{\rm IN})\!-\! \sum_{l=1}^{m}q_{l}^{\rm OUT}\epsilon(z_{l}^{\rm IN},\bar{z}_{l}^{\rm IN})  \Big]\langle {\rm OUT} |\mathcal{S}|{\rm IN} \rangle\, .  \nonumber
\end{eqnarray}

\begin{figure}[ht]
\centerline{ \includegraphics [width=0.8\columnwidth]{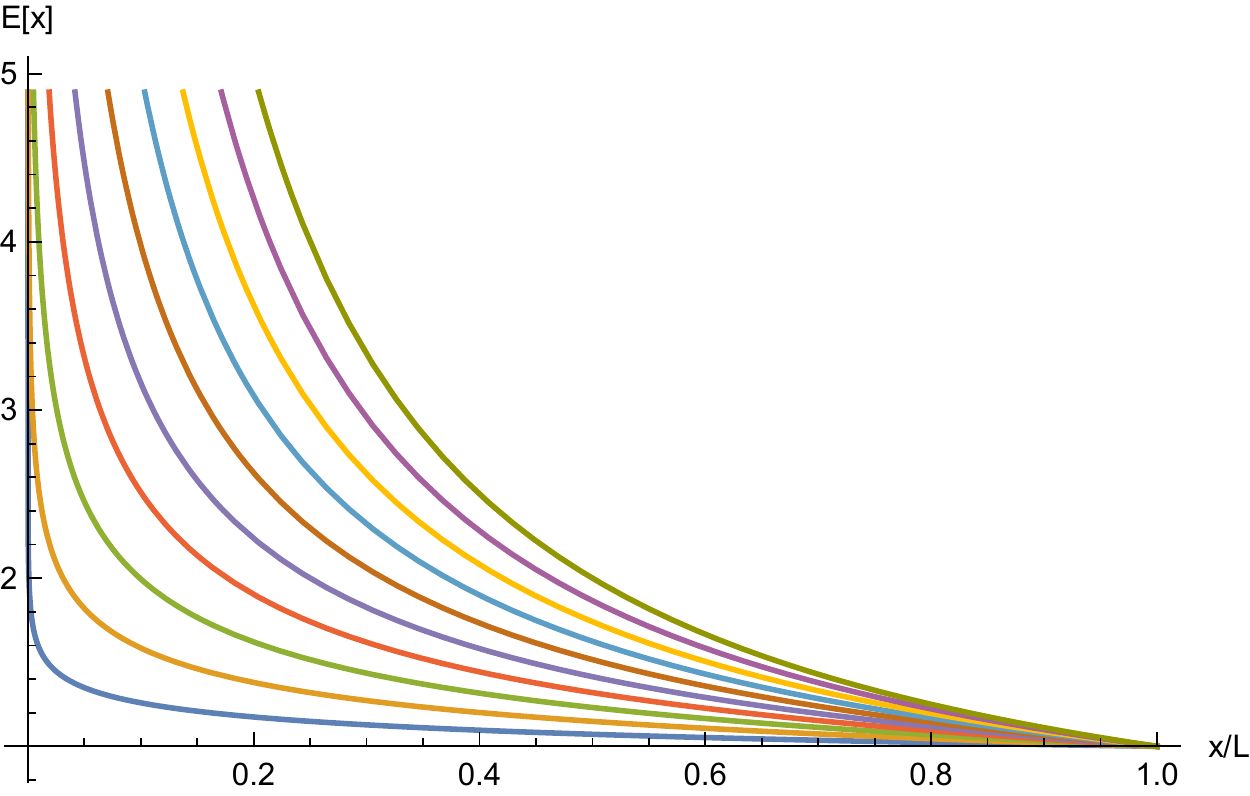}}
\caption{The predicted electric field due to the BMS analogous QED anomaly near the boundary, is displayed for different screening exponent parameters $\nu=0.1,\dots,1$ . 
The $x$ coordinate is normalized as the semi-metal length $L$, while $E$ is normalized to a typical scale $E_{0}$, dependent on the voltage applied. }
\end{figure}
As we previously commented, in vacuum the conformal invariant sphere casts
$$ds_{2}=2(1+z \bar{z})^{-2}dz d\bar{z}	\, .$$
The identity disconnected component of the Lorentz group is isomorphic to $PSL(2,\mathbb{C})$, and acts as a conformal transformation to the Riemann sphere. More specifically, the transformation has the form 
$$
z'=\frac{az+b}{cz+d}\,,
$$
where $a, b, c, d$ are complex number parameters constrained to $ad-bc=1$. The transformation then acts on the metric as 
$$
\frac{dz'd\bar{z}'}{(1+z' \bar{z}')^{2}}=\Omega^{2}(z,\bar{z})\frac{dz d\bar{z}}{(1+z\bar{z})^{2}}
\,,$$
where 
$$
\Omega(z,\bar{z})=\frac{1+z\bar{z}}{|az+b|^{2}+|cz+d|^{2}}\, . 
$$

The large asymptotic gauge symmetry $\mathcal{T}$ preserves the conformal invariance of the sphere, {\it i.e.} the algebra of the theory is $\mathcal{T}\times PSL(2,\mathbb{C})$. 

Such a group can be broken at the quantum level by conformal anomalies, induced from either gauge interactions or gravity, retaining the form
\begin{equation}
\label{kaak}
\mathcal{A}nomaly=\int_{M}\sqrt{g}[g^{\mu\nu}\langle T_{\mu\nu}\rangle-\langle g^{\mu\nu}T_{\mu\nu}\rangle]\, .
\end{equation}
For this class of QFT, $\mathcal{A}nomaly$ has the form
\begin{equation}
\label{jajalllk}
\mathcal{A}nomaly=\int_{M}\sqrt{g}[b_{1}F_{\mu\nu}F^{\mu\nu}+\mathcal{R}^{2}]\, ,
\end{equation}
where $\mathcal{R}^{2}$ here denotes 
$R^{2}$, 
$R_{\mu\nu}R^{\mu\nu}$,
$R_{\mu\nu\alpha\beta}R^{\mu\nu\alpha\beta}$, $\Box R$.

\vspace{2mm}
{\it Boundary QFT and induced anomaly.--- } Let us introduce a plane static boundary at a certain point $x$. Such a boundary induces a large asymptotic gauge symmetry anomaly. In  Boundary Quantum Field Theory \cite{Cardy:1984bb,McAvity:1993ue,McAvity:1995zd}, the renormalized current has a general structure close to the boundary $x\rightarrow 0$ of the form
\begin{equation}
\label{JRRR}
J_{R\mu}=\frac{1}{x^{3}}J_{\mu}^{(3)}+\frac{1}{x^{2}}J_{\mu}^{(2)}+\frac{1}{x}J_{\mu}^{(1)}+\log x J^{(0)}+...
\end{equation}
where the $J^{(n)}$ coefficients are dependent on the geometry. We focus on currents that are conserved once the anomaly term from Eq.(\ref{kaak}) is disregarded. 

Gauge invariance imposes the following constraints:
\begin{equation}
\label{JtJdJu}
J_{\mu}^{(3)}=J_{\mu}^{(2)}=0,
\end{equation}
\begin{equation}
\label{JtJdJud}
J_{\mu}^{(1)}=\alpha_{1}F_{\mu\nu}n^{\nu}+\alpha_{2}\mathcal{D}_{\mu}K+\alpha_{3}\mathcal{D}_{\nu}K^{\nu}_{\mu}+\alpha_{4}\star F_{\mu\nu} n^{\nu}\, , 
\end{equation}
where $F,n,\mathcal{D}, K,h$ are the background field gauge strenght,
the normal boundary vector, the covariant derivative and the extrinsic curvature induced on the boundary. We have expressed $n^{\mu}R_{\mu\nu}h_{\nu}^{\mu}$
by means of the extrinsic curvature, using the standard Gauss-Codazzi 
relation $n^{\mu}R_{\mu\nu}h^{\nu}_{\gamma}=\mathcal{D}_{\mu}K^{\mu}_{\gamma}-\mathcal{D}_{\gamma}K$.

The variation of the anomaly term Eq.~(\ref{kaak}) leads to the relation
with the integral on the bulk 
\begin{equation}
\label{deltaAe}
(\delta \mathcal{A})_{\partial M_{\epsilon}}=\Big(\int_{M}\sqrt{g}\delta A_{\mu}J_{R}^{\mu}\Big)_{\log \epsilon^{-1}}\, ,
\end{equation}
where $\epsilon$ is a x-regulator. Eq.~(\ref{deltaAe}) establishes a relation between the integrated anomaly $\mathcal{A}$ and a variation of the gauge field coupled to the $J^{R}$ current. The integral is UV logarithmically divergent. The metric can be rewritten in the Gauss normal coordinates
\begin{equation}
\label{dsdxh}
ds^{2}=dx^{2}+dy^{a} dy^{b}(h_{ab}-2rk_{ab}+r^{2}q_{ab}+...)\,,
\end{equation}
where $n_{\mu}=(1,0,0,0)$ is the normal vector. Let us chose a gauge $A_{r}=0$ and expand the gauge field around the boundary as $A_{i}=a_{i}+x_i A_{b}^{(1)}+...$.
The variation of the Weyl anomaly with respect to the gauge field 
leads to $(\delta \mathcal{A})_{\partial M}=4b_{1}\int_{\partial M}\sqrt{h}F^{b}_{n}\delta a_{b}$,
which is related to 
\begin{eqnarray}
\label{partM}
&\Big(  \int_{M}\sqrt{g}J^{\mu}_{R}\delta A_{\mu}\Big)_{\log 1/\epsilon}= \\
&\int_{\partial M}\sqrt{h}(\alpha_{1}F_{n}^{b}+\alpha_{2}\mathcal{D}^{b}k+\alpha_{3}\mathcal{D}_{j}k^{jb}
+\alpha_{4}\star F_{n}^{b}\big)\delta a_{b}\,.  \nonumber
\end{eqnarray}
For $\alpha_{1}=4b_{1} $ and $\alpha_{2,3,4}=0$, the latter leads to 
\begin{equation}
\label{Jbb}
J_{b}=\frac{4b_{1}F_{bn}}{x}\, ,
\end{equation}
which can also be rewritten as
\begin{equation}
\label{jak}
J^{\mu}=4b_{1}\frac{F^{\mu\nu}n_{\nu}}{x}\, . 
\end{equation}
The BMS analogous charges are shifted to 
\begin{equation}
Q^{+}_{\epsilon^{+}}\!+\!\Delta Q_{A}(r,\zeta,\bar{\zeta}) \!=\! Q^{+}_{\epsilon^{+}} \!+\!\int_{\mathcal{I}_{+}} \!\!\!\! du \,d^{2}\zeta\, \epsilon\, \gamma_{\zeta\bar{\zeta}}\, 4b_{1}\,\frac{F^{bn}}{x(r,\zeta,\bar{\zeta})}]\,, \nonumber
\end{equation}
where $x(r,z,\bar{z})$ is understood as a mapping between standard Minkowski in Cartesian-coordinates and the Penrose diagram coordinates. This result is somehow expected, since the new charge distribution on the x-axis does not allow to define conserved Noether charges on the asymptotic null infinity. Indeed, the phenomenon is related to a change of the vector and scalar potential in the asymptotic limit. 

{\it Electric field from the anomaly, and suggested experiment.--- } The new anomaly percolates on a new equation for the electric field
\begin{equation}
\label{ajka}
\partial_{x}E_{x}=-\frac{1}{\epsilon_{0}}\frac{2\beta_{e}}{ec\hbar}\frac{E_{x}(x)}{x},\qquad x>0\, , 
\end{equation}
which leads to an electrostatic potential and field, the profile of which is
\begin{equation}
\phi(x)=\phi_{0}-\frac{Cx^{1-\nu}}{1-\nu}\, ,  \qquad
E_{x}=\frac{C}{x^{\nu}}\, ,
\end{equation}
where $\nu$ is the critical anomalous exponent, which is related to $\beta_q$, the renormalization $\beta$ function of the electric coupling $q$, and the vacuum dielectric constant $\epsilon_{0}$ by 
\begin{eqnarray}
\label{jakka}
\nu=\frac{2\beta_{q}}{q c \hbar \epsilon_{0}}\, . 
\end{eqnarray}
If the presence of the boundary is within a dielectric material, Eq.~\eqref{jakka} can be generalized with the dielectric of the material by substituting $\epsilon_{0}\rightarrow \epsilon$. In Fig.2 the predicted electric field corrected by the BMS analogous anomaly is displayed.

As a simple experimental set-up can be envisaged to test the effect of the quantum anomaly of analogous BMS symmetries arising from the boundary. We can propose exactly the same concept of Refs.~\cite{Arjona:2019lxz,Chernodub:2019blw}, being sufficient to consider a Dirac or Weyl semi-metal material, applying a difference of potential on the two sides of it. In this way, we can test the electrostatic profile $\phi(x)$, which is affected by the quantum screening effect.  
Indeed, the conformal anomaly pointed out in Refs.~\cite{Arjona:2019lxz,Chernodub:2019blw} is related to the BMS analogue in QED. However, there is a new prediction of BMS anomaly that cannot be captured by the conformal anomaly. This new class of effects are related to the electromagnetic radiation that may be detected from these materials, providing a new memory effect as for BMS and gravity. As a matter of fact, the soft Bremsstrahlung effect is described by the QED soft limits, in turn related to the Large Asymptotic Symmetry. An anomaly of it percolates onto the modification of the soft Bremsstrahlung effect. Indeed, being Eq.~\eqref{ajak} anomalous, the asymptotic future and past charges in Eqs.~\eqref{Qaaa}-\eqref{kaka} are modified by the anomaly. This affects any QED scattering process considered in the material. 
These aspects certainly deserve future extended experimental and theoretical analysis 
beyond the purposes of this paper.

\vspace{0.1cm}
\noindent 
{\it Conclusions.---} The infinite dimensional asymptotic gauge symmetry in QED is quantum anomalous in presence of a magnetized boundary. This effect may be tested in Dirac or Weyl semimetals over the next future. Thus, it might provide a better understanding of the role of infinite dimensional symmetries in quantum physics, including the BMS symmetries in gravity, which allow a memory effect for quantum black hole physics.


\acknowledgements 
\noindent
We wish to acknowledge support by the NSFC, through the grant No.~11875113, by the Shanghai Municipality, through the grant No.~KBH1512299, and by Fudan University, through the grant No.~JJH1512105.


\end{document}